\documentclass[conference]{IEEEtran}

\usepackage[final]{graphicx}
\usepackage{booktabs}
\usepackage{array}
\usepackage{url}

\newcommand{\detected}{D}
\newcommand{\missed}{M}
\newcommand{\na}{N/A}

\begin{document}

\title{PDFa11yMut: Mutation-Based Evaluation of Automated PDF Accessibility Validators}

\author{
\IEEEauthorblockN{Gauri Jain}
\IEEEauthorblockA{University of California, Santa Cruz\\
Email: gjain1@ucsc.edu}
}

\maketitle

\begin{abstract}
\normalfont
Automated validators are widely used to assess PDF accessibility, but their detection boundaries are difficult to measure without controlled ground truth. A validator-clean PDF may be free of the implemented machine-checkable failures, or it may contain accessibility-relevant structural defects that the validator does not detect. This paper introduces PDFa11yMut, a controlled mutation-based benchmark for evaluating automated PDF accessibility validators against independently verified PDF structure defects.

We generated 23 verified mutants from five validator-clean golden PDFs using five mutation operators targeting reading order, omitted assistive structure, heading hierarchy, marked-content association, and internal marked-content sequence. Each generated mutant was independently verified at the PDF structure level and checked for parseability, page-count preservation, and absence of detected visual differences. We then evaluated the mutants with three validators: PAC, Adobe Acrobat Accessibility Checker, and veraPDF, with veraPDF run using the PDF/UA-1 profile.

The validators detected a small and highly operator-specific subset of the verified mutants. PAC detected 4/23 mutants (17.4\%), while Acrobat and veraPDF each detected 5/23 mutants (21.7\%). All three validators detected every generated heading-hierarchy mutant. None detected the reading-order swap, omitted-structure, or internal MCID-order mutants. Acrobat and veraPDF also detected one marked-content association mutant that PAC missed. The practical implication is that a validator-clean result can coexist with verified, visually invisible structure defects that affect the assistive representation of a PDF.
\end{abstract}


\section{Introduction}

PDF accessibility validators are software systems that check whether document structures satisfy accessibility-related requirements. Their output is often treated as evidence that a PDF is accessible or inaccessible, but this interpretation is only as strong as the checker rules being exercised. This matters because PDFs remain a widely used global format for education, research, government, healthcare, legal records, financial documents, and public services. When a visually correct PDF exposes headings, lists, tables, or reading order incorrectly to assistive technologies, the barrier is not visible on the page but can still affect access to essential information. Some of these defects may fall outside a validator's automated checks.

This creates a benchmark problem. A validator-clean result may mean that the PDF is structurally sound with respect to the implemented checks. It may also mean that the PDF contains an accessibility-relevant defect outside the validator's automated detection boundary. Without controlled ground truth, it is difficult to tell whether a checker is detecting a defect, missing it, or simply not designed to evaluate that property.

Mutation testing provides a way to make this boundary measurable. Rather than relying only on naturally occurring PDFs with ambiguous defects, a mutation-based benchmark starts from validator-clean golden documents, injects controlled defects, verifies those defects independently, and measures whether validators detect them. Ma11y applies this idea to web accessibility testing tools \cite{tafreshipour2024ma11y}, and GateTruth uses mutation testing as a benchmark-auditing method in hardware design evaluation \cite{bhadra2026gatetruth}. PDF accessibility lacks an equivalent mutation benchmark for automated validators.

This paper presents PDFa11yMut, a controlled mutation benchmark for automated PDF accessibility validators. PDFa11yMut starts from validator-clean golden PDFs, injects structure-level PDF accessibility mutations, verifies the intended structural change independently from validator output, and then measures whether validators report a relevant automated failure. This design turns an otherwise vague concern about checker limitations into measurable evidence: for each injected defect, the benchmark records whether the defect is present, whether the page still renders as expected, and whether the validator reports a corresponding failure. A surviving mutant is therefore interpreted carefully: not as proof that the PDF is accessible, and not as proof that a validator is generally inaccurate, but as evidence that a verified mutation survived the validator's automated checks under the tested configuration.

The paper makes the following contributions:
\begin{itemize}
    \item a controlled mutation methodology for PDF accessibility structures;
    \item five mutation operators targeting accessibility-relevant PDF structure defects;
    \item a verified mutant set derived from five golden PDFs;
    \item a three-validator detection matrix for PAC, Adobe Acrobat Accessibility Checker, and veraPDF; and
    \item an analysis showing operator-specific detection behavior and one cross-validator disagreement case.
\end{itemize}

\section{Research Questions}

The study addresses four research questions:

\textbf{RQ1.} To what extent do automated PDF accessibility validators detect independently verified structure-only accessibility mutations in otherwise validator-clean PDF/UA-oriented documents?

\textbf{RQ2.} Are detection outcomes concentrated in particular mutation operators rather than distributed evenly across reading-order, omission, list/table association, and MCID-order defects?

\textbf{RQ3.} How much do commonly used validators agree or disagree on the same controlled PDF accessibility mutants, and which mutation cases explain disagreement?

\textbf{RQ4.} What does mutation-based evaluation reveal about the boundary between machine-checkable conformance failures and verified accessibility-relevant defects that survive automated checking?

\section{Background and Related Work}

\subsection{PDF Accessibility Evaluation}

Recent work has shown that PDF accessibility remains a persistent problem in scholarly communication. Kumar and Wang studied approximately 20,000 scholarly PDFs published between 2014 and 2023 and found low compliance across measured accessibility criteria, including a reported decline after 2019 \cite{kumar2024crisis}. Their work demonstrates the scale of the problem and the practical role of automated accessibility checking in large document collections. It also cautions against treating automated checker output as a complete account of user-facing accessibility.

This recent work builds on earlier studies of scholarly PDF accessibility. Brady, Zhong, and Bigham analyzed conference proceedings from CHI, ASSETS, and W4A and documented accessibility barriers in research-paper PDFs \cite{brady2015pdf}. Nganji later examined PDFs from disability-related journals and found low rates of tagging and alternative text despite the availability of HTML alternatives \cite{nganji2018pdf}. Ribera, Pozzobon, and Sayago reported a case study on producing accessible proceedings for DSAI 2016, emphasizing the labor and workflow coordination required to produce accessible PDFs \cite{ribera2020proceedings}. Together, these studies show that PDF accessibility failures are persistent across venues and that tool support remains central to scalable evaluation and remediation.

Kumar, Padath, and Wang later proposed a benchmark for PDF accessibility evaluation that includes expert-validated annotations across seven criteria, including alternative text quality, logical reading order, semantic tagging, table structure, functional hyperlinks, color contrast, and font readability \cite{kumar2025benchmarking}. Their benchmark evaluates both automated and LLM-based approaches and emphasizes that PDF accessibility assessment requires more nuanced labels than a simple binary pass/fail result. PDFa11yMut is complementary: expert-labeled datasets measure evaluator performance against human judgments, while mutation benchmarks isolate whether a checker detects specific injected defects.

Work on PDF remediation tools also motivates better evaluator diagnostics. Paliwal et al. present FormA11y, a tool for remediating PDF forms for accessibility \cite{paliwal2024forma11y}. Remediation work highlights the complexity of repairing accessibility defects once they exist. PDFa11yMut addresses an adjacent problem: before remediation, tool builders and researchers need benchmark data showing which classes of PDF structure defects automated evaluators can detect reliably.

\subsection{PDF/UA, Matterhorn, and Machine-Checkable Criteria}

PDF/UA provides technical requirements for accessible PDF files, while the Matterhorn Protocol operationalizes PDF/UA-1 conformance testing as a set of checkpoints and failure conditions \cite{pdfa2021matterhorn}. This distinction matters because the protocol separates checks that can be determined by software from checks that require human judgment. The PDF Association describes the Matterhorn Protocol as a list of possible ways to fail PDF/UA-1, and veraPDF's documentation similarly notes that, for PDF/UA, veraPDF performs only machine-verifiable checks \cite{verapdfvalidation}.

This machine/human boundary is central to PDFa11yMut. Some properties, such as malformed heading hierarchy or invalid list containment, are well suited to rule-based validation. Others, such as meaningful reading order, the usefulness of alternative text, or whether a structural association matches the author's intended semantics, may require contextual reasoning or human judgment. Adobe's Acrobat checker documentation also reflects this hybrid model: it describes checks that pass or fail, automatic fixes for some checks, and instructions for items requiring manual fixes \cite{adobeaccessibility}.

PDFa11yMut does not assume that a missed mutant proves that a validator is incorrect. Instead, it records whether a verified accessibility-relevant structural mutation survives the validator's automated checks. Automated checker output is useful evidence, but it should not be treated as a complete accessibility oracle.

\subsection{Accessibility Mutation Testing}

Mutation testing has recently been applied to accessibility evaluation outside the PDF domain. Tafreshipour et al. introduced Ma11y, a mutation framework for web accessibility testing \cite{tafreshipour2024ma11y}. Ma11y defines 25 mutation operators that intentionally violate accessibility principles and uses an automated oracle to determine whether web accessibility testing tools detect the injected mutants. Their evaluation on real-world websites found that current tools missed nearly half of the injected accessibility bugs.

Ma11y is the most direct methodological predecessor to PDFa11yMut. Both projects use mutation analysis to evaluate accessibility testing tools against controlled defects rather than relying only on naturally occurring accessibility issues. PDFa11yMut differs in domain and technical substrate. Ma11y mutates web content, where accessibility is typically represented through HTML, ARIA, CSS, and DOM-level properties. PDFa11yMut mutates PDF structure trees, marked-content identifiers, heading tags, list/table structures, and marked-content associations.

Accessibility mutation testing has also been explored for mobile applications. Silva, Vergilio, and Endo proposed mutation testing for Android accessibility, defining operators derived from accessibility faults and using mutants to assess and improve test suites \cite{silva2022android}. This work reinforces the broader idea that accessibility-specific mutation operators can reveal weaknesses in existing testing workflows. PDFa11yMut extends that principle to static document accessibility and PDF/UA-oriented validation.

\subsection{Mutation Testing for Benchmark Auditing}

GateTruth applies mutation testing to register-transfer-level design benchmarks for large language models \cite{bhadra2026gatetruth}, demonstrating how controlled mutations can be used to audit whether an evaluation framework reliably distinguishes valid artifacts from defective ones. Its central methodological principle is that evaluator reliability should itself be tested against controlled, meaningful defects before benchmark pass rates are treated as strong evidence. PDFa11yMut applies this benchmark-auditing principle to PDF accessibility validation. Rather than assuming that validator-clean output demonstrates the absence of accessibility-relevant structural defects, PDFa11yMut introduces independently verified mutations and measures whether they survive automated validation. In this way, mutation testing provides a systematic mechanism for characterizing the detection boundaries of PDF accessibility validators.

\section{Methodology}

\subsection{Corpus}

The experiment began with five golden PDFs, labeled G01 through G05. These PDFs served as the baseline corpus from which mutants were generated. Before interpreting mutation results, the golden PDFs were checked with the evaluated validators to establish that validator-reported failures in mutants were not inherited from the baseline documents.

For veraPDF, all five golden PDFs passed the PDF/UA-1 validation profile. The veraPDF report identifies version 1.30.2, the GreenField parser, and a PDF/UA-1 validation profile containing 106 rules. The recorded Acrobat baseline reported 29 passed checks, 0 failed checks, 2 manual checks, and 1 skipped check for each golden PDF. The two Acrobat manual-check prompts were reviewed manually and did not identify baseline defects. The recorded PAC baseline also reported the golden PDFs as passing. The Acrobat manual-check items were documented separately from automated mutation-detection outcomes.

\subsection{Mutation Operators}

PDFa11yMut uses five mutation operators. Each operator targets the assistive representation of the PDF while preserving visible content where possible. Table~\ref{tab:operators} summarizes the operators.

\begin{table*}[t]
\caption{PDFa11yMut mutation operators and intended accessibility effects.}
\label{tab:operators}
\centering
\small
\begin{tabular}{p{0.06\textwidth}p{0.19\textwidth}p{0.34\textwidth}p{0.33\textwidth}}
\toprule
ID & Operator & Structural change & Intended accessibility effect \\
\midrule
M01 & Structural reading-order swap & Swap adjacent structural children, such as list items or table rows. & Assistive technology encounters content in a different order from the intended visible order. \\
M02 & Omission from assistive structure & Remove a meaningful visible element from the structure tree while leaving page content visible. & A visible heading or content item is absent from the assistive representation. \\
M03 & Incorrect heading level & Change a heading tag to an inappropriate lower level. & The document exposes a skipped or misleading heading hierarchy. \\
M04 & Wrong marked-content association & Preserve structural sibling order while swapping underlying marked-content associations. & A tag position points to the wrong visible content. \\
M05 & Internal MCID sequence reversal & Reverse the internal marked-content sequence inside one structural element. & A list item or paragraph exposes its internal chunks in the wrong order. \\
\bottomrule
\end{tabular}
\end{table*}

The practical impact of each operator comes from the difference between the
visual page and the structure exposed to assistive technologies. M01 models a
case where the page looks correct but the assistive reading sequence is wrong:
two list items that visually appear as ``submit the form'' and then ``wait for
confirmation'' may be exposed in the opposite order. M02 models visible content
that remains on the page but is absent from the tagged representation, such as
a visible heading that is no longer exposed through the document structure.
M03 models a misleading document outline, such as jumping from H1 to H3 without
an intervening H2. M04 models a tag that occupies the expected structural
position but points to the wrong visible content; for example, the first list
item's structural tag may reference the text belonging to the second list item,
while the visual page remains unchanged. M05 models a single structural element
whose internal content chunks are exposed in the wrong sequence; for example,
multiple marked-content chunks belonging to one paragraph or list item may be
encountered in reverse order by software consuming the tagged structure.

\subsection{Mutant Generation and Verification}

\begin{figure}[t]
\centering
\includegraphics[width=\linewidth]{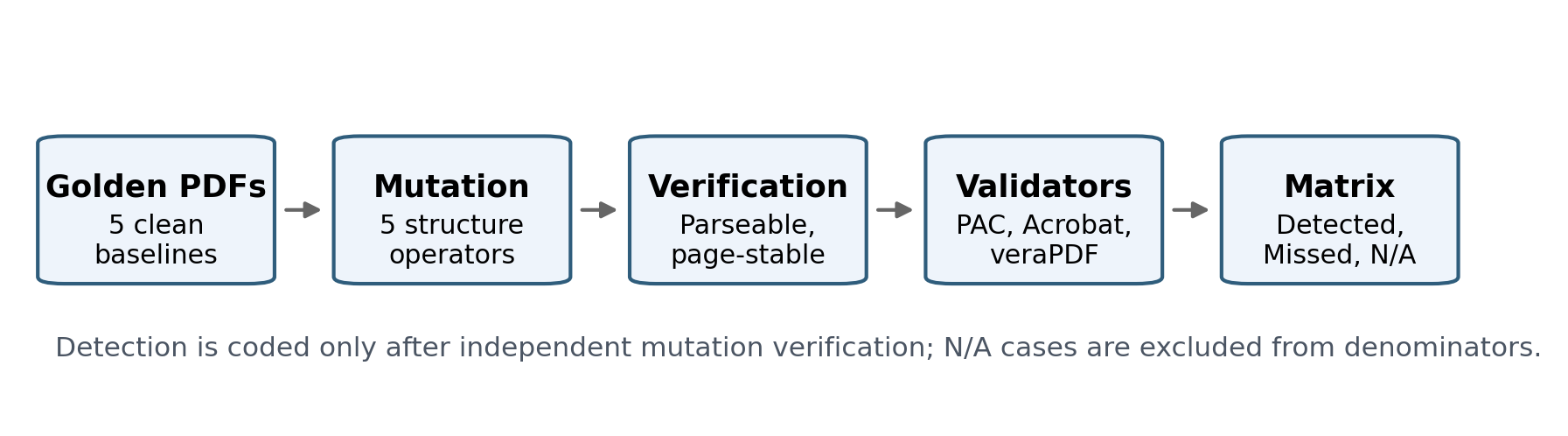}
\caption{PDFa11yMut experimental pipeline. Detection is coded only after independent mutation verification.}
\label{fig:pipeline}
\end{figure}

Figure~\ref{fig:pipeline} summarizes the experimental pipeline. The workflow selected a suitable target object in a golden PDF, applied a mutation operator, parsed the resulting PDF, verified that page count was unchanged, compared the targeted structure before and after mutation, checked that the intended mutation was present, checked for unexpected structural changes and detected visual differences, ran each validator on the verified mutant, and coded the validator result as Detected, Missed, or N/A.

The five operators were applied where suitable targets existed in each golden PDF. This produced 23 generated mutants. Two potential cases were not generated and are preserved as N/A rather than counted as missed or detected: G05-M03 was not generated because G05 did not contain a suitable heading-hierarchy target, and G05-M05 was not generated because G05 did not contain a suitable multi-MCID target. Preserving these N/A cases avoids inflating the denominator with mutants that were not generated.

Validator detection was separated from mutation verification. A mutant was included in the detection denominator only if the mutation generation and verification pipeline confirmed generation success, parseability, unchanged page count, mutation verification, no unexpected structural changes, and no detected visual difference. A validator result of ``missed'' means that the validator did not report a relevant automated failure for a mutant whose intended defect was independently verified. It does not mean the PDF is accessible, and it does not measure the validator's general accuracy.

\subsection{Validators and Detection Coding}

The experiment evaluated PAC, Adobe Acrobat Accessibility Checker, and veraPDF. PAC evidence identifies PAC 26.1. The recorded PAC experiment found that all golden PDFs passed. Among mutants, PAC reported Structure Elements failures for G01-M03, G02-M03, G03-M03, and G04-M03. All other generated mutants were recorded as passing.

Adobe Acrobat Continuous Release version 2026.001.21789 (64-bit) was used as a widely available commercial PDF accessibility checker. The recorded Acrobat results found no automated failures in the five golden PDFs. Among mutants, Acrobat detected G01-M03, G02-M03, G03-M03, G04-M03, and G02-M04. For the M03 mutants, Acrobat reported heading nesting failures. For G02-M04, Acrobat reported list-related failures.

veraPDF was used as an open-source standards-oriented validator with the PDF/UA-1 validation profile. The veraPDF report used version 1.30.2 and the GreenField parser. veraPDF detected G01-M03, G02-M03, G03-M03, G04-M03, and G02-M04. The M03 mutants failed ISO 14289-1:2014 Clause 7.4.2 Test 1. G02-M04 failed four list-structure checks under ISO 14289-1:2014 Clause 7.2.

No validator repair, auto-tagging, or manual remediation was applied to the PDFs before recording results. Manual-check prompts were not coded as automated detections. For example, Acrobat's recurring manual-check items on the golden PDFs were manually reviewed and recorded as passing, but they were not counted as mutation detections because the experiment measures automated validator detection under fixed checker settings. A validator failure was counted as a detection only when the failed rule aligned with the mutation's intended defect category.

\section{Results}

\subsection{Overall Detection Rates}

\begin{figure}[t]
\centering
\includegraphics[width=\linewidth]{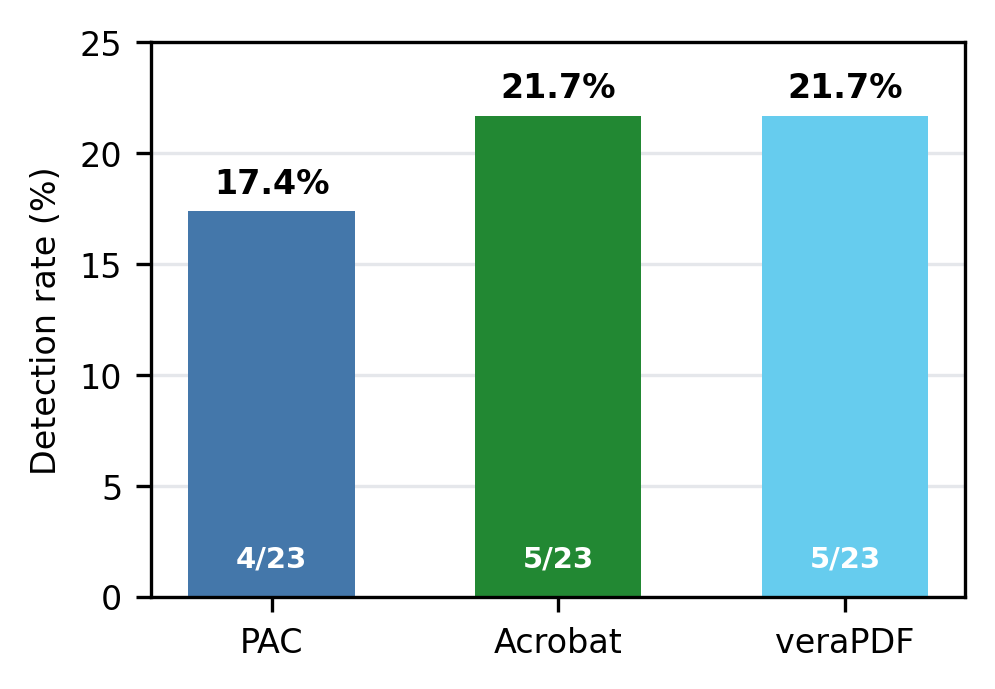}
\caption{Overall mutation-detection rates across 23 generated and independently verified mutants.}
\label{fig:overall}
\end{figure}

Across 23 generated and independently verified mutants, PAC detected 4 mutants, Acrobat detected 5 mutants, and veraPDF detected 5 mutants. Table~\ref{tab:overall} and Fig.~\ref{fig:overall} summarize the overall mutation-detection rates. These values are mutation-detection rates for this controlled benchmark; they should not be read as general accessibility accuracy scores.

\begin{table}[t]
\caption{Overall mutation-detection rates.}
\label{tab:overall}
\centering
\begin{tabular}{lrrrr}
\toprule
Validator & Detected & Missed & Total & Rate \\
\midrule
PAC & 4 & 19 & 23 & 17.4\% \\
Acrobat & 5 & 18 & 23 & 21.7\% \\
veraPDF & 5 & 18 & 23 & 21.7\% \\
\bottomrule
\end{tabular}
\end{table}

This matters because most verified structural defects in the benchmark survived all evaluated automated checks. In a workflow that relies only on these automated outputs, the surviving mutants would remain indistinguishable from the validator-clean baselines unless a separate structural inspection, assistive-technology check, or targeted structural test were used.

\subsection{Detection by Mutation Operator}

Detection was highly concentrated in one mutation operator. All three validators detected every generated M03 heading-hierarchy mutant, giving each validator a 4/4 detection rate for M03. No validator detected any M01, M02, or M05 mutant. These operators covered structural reading-order swaps, omitted assistive structure, and internal MCID sequence reversal. For M04, Acrobat and veraPDF each detected 1/5 mutants, while PAC detected 0/5. The only detected M04 case was G02-M04.

\begin{figure}[t]
\centering
\includegraphics[width=\linewidth]{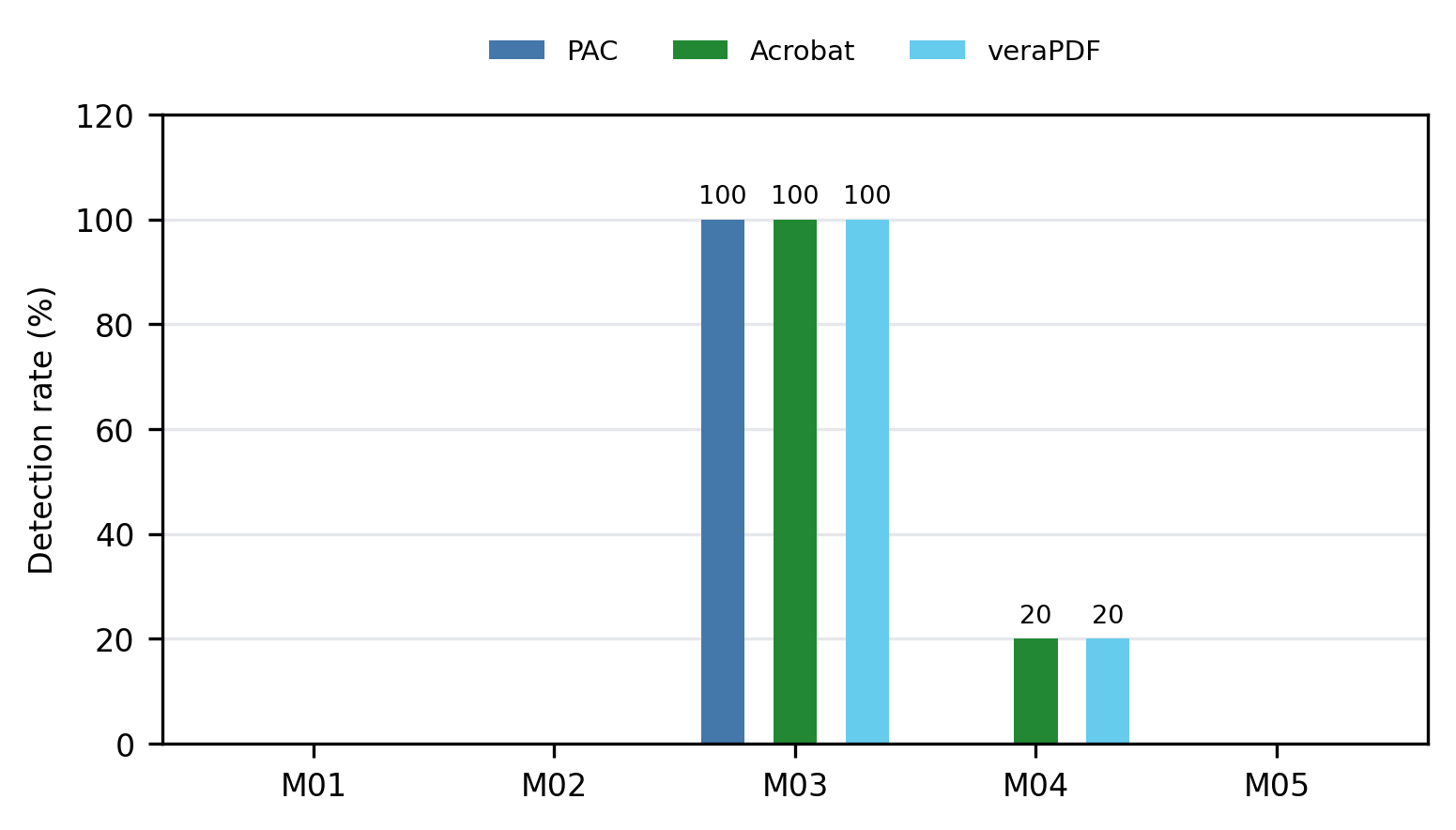}
\caption{Detection by mutation operator. Detection is concentrated in M03; Acrobat and veraPDF also detect one M04 mutant.}
\label{fig:peroperator}
\end{figure}

Table~\ref{tab:peroperator} and Fig.~\ref{fig:peroperator} summarize the operator-specific pattern.

\begin{table}[t]
\caption{Detection by mutation operator.}
\label{tab:peroperator}
\centering
\small
\begin{tabular}{llrrr}
\toprule
ID & Operator & PAC & Acrobat & veraPDF \\
\midrule
M01 & Reading order & 0/5 & 0/5 & 0/5 \\
M02 & Omitted structure & 0/5 & 0/5 & 0/5 \\
M03 & Heading level & 4/4 & 4/4 & 4/4 \\
M04 & MC association & 0/5 & 1/5 & 1/5 \\
M05 & MCID order & 0/4 & 0/4 & 0/4 \\
\bottomrule
\end{tabular}
\end{table}

These misses are practically meaningful because the affected properties influence what assistive technologies can navigate or read, even when the page appearance is preserved. For example, omitted structure can remove visible content from the tagged representation, and reading-order or MCID-order changes can alter the sequence exposed to nonvisual users.

\subsection{Validator Agreement}

Acrobat and veraPDF agreed on all 23 generated mutants at the detected/missed level. PAC agreed with Acrobat and veraPDF on 22 of 23 generated mutants. The only disagreement case was G02-M04, which PAC missed and Acrobat and veraPDF detected.

\begin{table}[t]
\caption{Validator agreement on generated mutants.}
\label{tab:agreement}
\centering
\small
\begin{tabular}{lrrrl}
\toprule
Pair & Agree & Disagree & Total & Case \\
\midrule
PAC--Acrobat & 22 & 1 & 23 & G02-M04 \\
PAC--veraPDF & 22 & 1 & 23 & G02-M04 \\
Acrobat--veraPDF & 23 & 0 & 23 & None \\
\bottomrule
\end{tabular}
\end{table}

G02-M04 is the clearest disagreement case in the experiment. The mutation verification data confirmed that the list order remained structurally stable while the first two list items pointed to each other's underlying content. PAC reported no automated failure for this mutant. Acrobat detected G02-M04 and reported two list-related failures: List items and Lbl/LBody. veraPDF also detected G02-M04 and reported four failures under ISO 14289-1:2014 Clause 7.2, covering list containment and allowed-child constraints.

\begin{table*}[t]
\caption{Final detection matrix. D = Detected, M = Missed, N/A = not generated.}
\label{tab:matrix}
\centering
\small
\begin{tabular}{llccc|llccc|llccc}
\toprule
Mutant & Op. & PAC & Acro. & vera & Mutant & Op. & PAC & Acro. & vera & Mutant & Op. & PAC & Acro. & vera \\
\midrule
G01-M01 & M01 & \missed & \missed & \missed & G02-M04 & M04 & \missed & \detected & \detected & G04-M02 & M02 & \missed & \missed & \missed \\
G01-M02 & M02 & \missed & \missed & \missed & G02-M05 & M05 & \missed & \missed & \missed & G04-M03 & M03 & \detected & \detected & \detected \\
G01-M03 & M03 & \detected & \detected & \detected & G03-M01 & M01 & \missed & \missed & \missed & G04-M04 & M04 & \missed & \missed & \missed \\
G01-M04 & M04 & \missed & \missed & \missed & G03-M02 & M02 & \missed & \missed & \missed & G04-M05 & M05 & \missed & \missed & \missed \\
G01-M05 & M05 & \missed & \missed & \missed & G03-M03 & M03 & \detected & \detected & \detected & G05-M01 & M01 & \missed & \missed & \missed \\
G02-M01 & M01 & \missed & \missed & \missed & G03-M04 & M04 & \missed & \missed & \missed & G05-M02 & M02 & \missed & \missed & \missed \\
G02-M02 & M02 & \missed & \missed & \missed & G03-M05 & M05 & \missed & \missed & \missed & G05-M03 & M03 & \na & \na & \na \\
G02-M03 & M03 & \detected & \detected & \detected & G04-M01 & M01 & \missed & \missed & \missed & G05-M04 & M04 & \missed & \missed & \missed \\
 & & & & & & & & & & G05-M05 & M05 & \na & \na & \na \\
\bottomrule
\end{tabular}
\end{table*}

\section{Discussion}

\subsection{Mutation Detection Is Operator-Specific}

The strongest result is not simply that overall detection rates were low. The more important finding is that detection was concentrated by operator. All validators detected M03 heading-hierarchy violations, but none detected M01, M02, or M05, and only Acrobat and veraPDF detected one M04 case. This suggests that automated PDF accessibility validators may be highly sensitive to certain standards-checkable structural constraints while remaining insensitive to other accessibility-relevant structure changes. This is consistent with the distinction between machine-checkable conformance rules and defects that require deeper semantic interpretation.

\subsection{Surviving Mutants Are Not Proof of Accessibility}

The 18 mutants missed by all three validators were not treated as valid or accessible PDFs. They were treated as verified mutants that survived the evaluated automated checks. This distinction is essential for avoiding overstatement. For example, a structural reading-order swap can be accessibility-relevant even if it does not violate a validator's automated rule. A validator may correctly decline to flag a problem that requires semantic judgment. The benchmark therefore measures whether a validator detects specific controlled mutations, not whether the validator makes a complete accessibility judgment.

\subsection{G02-M04 Shows Cross-Validator Sensitivity Differences}

G02-M04 demonstrates that validators may differ even within the same operator class. PAC missed all M04 mutants, while Acrobat and veraPDF detected G02-M04. Because Acrobat and veraPDF both identified list-structure problems, this case provides stronger evidence than a generic pass/fail disagreement. At the same time, Acrobat and veraPDF did not detect the other M04 mutants. This suggests that detection may depend not only on the abstract mutation operator but also on the specific local PDF structure where the mutation is applied.

\subsection{Implications for Validator Use}

The results support a practical distinction between using validators as conformance aids and treating them as complete accessibility oracles. PAC, Acrobat, and veraPDF were effective on the generated heading-hierarchy defects, and Acrobat and veraPDF also detected one list-related association defect. However, the same tools did not report relevant automated failures for verified mutations that changed reading sequence, omitted tagged content, or reversed internal marked-content order. For document producers, this means validator-clean output should be understood as evidence about the implemented automated checks, not as evidence that the assistive representation is free of structure-level defects. For tool builders, the surviving mutants identify concrete regression tests for future checker rules.

\section{Threats to Validity}

\textbf{Corpus size.} The corpus contains five golden PDFs and 23 generated mutants. The corpus supports controlled within-benchmark claims about these five PDFs and 23 verified mutants. Broader generalization across PDF genres, authoring pipelines, and defect types requires additional corpus expansion.

\textbf{Mutation operator coverage.} The five mutation operators target structure-level defects. They do not cover all PDF accessibility problems, such as alternative text quality, table header semantics, language metadata, color contrast, form labeling, or natural-language reading-order coherence.

\textit{Validator configuration.} The results depend on the validator versions and settings used in the experiment. Different versions or configurations may report different outcomes. The evaluated versions were PAC 26.1, Adobe Acrobat Continuous Release 2026.001.21789 (64-bit), and veraPDF 1.30.2.

\textbf{Detection coding.} The study codes detection based on relevant automated validator failures. Some validator outputs may include warnings, manual-check prompts, or failures that do not correspond directly to the injected mutation. This paper treats only relevant automated failures as detections.

\textbf{Independent verification scope.} The mutation verification process confirms the structural presence of the injected defects and the visual-preservation properties needed for this benchmark. User studies and assistive-technology traces would address a different question: how each verified structural defect affects end-user experience.

\section{Limitations and Future Work}

This study establishes an initial mutation-based methodology for evaluating
automated PDF accessibility validators using independently verified structural
defects. The present benchmark focuses on five structure-level mutation
operators applied across five golden PDFs, enabling controlled comparison of
validator behavior while preserving independently verified ground truth.
Future work can build on this methodology by introducing additional mutation
operators and document structures to investigate whether the operator-specific
detection patterns observed here generalize across broader PDF accessibility
scenarios.

The public PDFa11yMut repository provides machine-readable analysis artifacts
and benchmark scripts to support reproduction and extension of the reported
results. Future work can extend PDFa11yMut by evaluating additional validators,
studying the effects of surviving mutants on assistive-technology output, and
examining how mutation-based benchmarks can inform validator regression testing
and remediation workflows.

\section{Artifact Availability}

The PDFa11yMut research artifact is publicly available at
\url{https://github.com/gaurijain21/pdfa11ymut}.
The repository contains the paper source and figures, machine-readable
analysis data, the validator detection matrix, the analysis workbook,
and scripts used to generate the experimental analysis and paper
figures. The released data preserve N/A cases explicitly and report
per-mutant validator outcomes, per-operator detection rates, overall
detection rates, and validator agreement results.

The current public release provides the reproducible analysis artifacts
and benchmark scripts associated with the results reported in this paper.
The golden PDFs, generated mutant PDFs, and raw validator reports are not
included in the public release while redistribution rights and potential
identifying or private information are reviewed.

\section{Conclusion}

PDFa11yMut demonstrates that controlled mutation testing can reveal how automated PDF accessibility validators respond to independently verified structural defects. In this benchmark, three validators detected all generated heading-hierarchy mutants but missed most other verified mutations. Acrobat and veraPDF detected one list-related marked-content association mutant that PAC missed, producing the only cross-validator disagreement case.

The contribution is both methodological and practical. Methodologically, PDFa11yMut separates mutation verification from validator output, giving researchers a way to test checker behavior against known structural defects rather than ambiguous naturally occurring PDFs. Practically, the results show that validator-clean outcomes can coexist with verified defects in the assistive structure, including defects that preserve visual rendering. These results do not show that any validator is generally accurate or inaccurate. They show that mutation-based evaluation can make validator behavior more observable, especially at the boundary between machine-checkable PDF/UA conformance rules and accessibility-relevant defects that survive automated checking. PDFa11yMut therefore provides a reproducible basis for studying PDF accessibility validator sensitivity and for building regression tests that target validator blind spots.


\begin{thebibliography}{00}

\bibitem{adobeaccessibility}
Adobe, ``Acrobat Enterprise Toolkit: Accessibility Checker Preferences,'' 2026. [Online]. Available: \url{https://www.adobe.com/devnet-docs/acrobatetk/tools/PrefRef/Windows/Accessibility.html}

\bibitem{bhadra2026gatetruth}
M. Bhadra, ``GateTruth: Auditing the Rigor of RTL Design Benchmarks via Mutation Testing,'' arXiv:2608.12635, 2026. [Online]. Available: \url{https://arxiv.org/abs/2608.12635}

\bibitem{brady2015pdf}
E. Brady, Y. Zhong, and J. P. Bigham, ``Creating accessible PDFs for conference proceedings,'' in \emph{Proc. 12th Web for All Conf.}, 2015, doi: 10.1145/2745555.2746665.

\bibitem{kumar2025benchmarking}
A. Kumar, T. Padath, and L. L. Wang, ``Benchmarking PDF Accessibility Evaluation: A Dataset and Framework for Assessing Automated and LLM-Based Approaches for Accessibility Testing,'' in \emph{Proc. 27th Int. ACM SIGACCESS Conf. Computers and Accessibility}, 2025, doi: 10.1145/3663547.3746380.

\bibitem{kumar2024crisis}
A. Kumar and L. L. Wang, ``Uncovering the New Accessibility Crisis in Scholarly PDFs,'' in \emph{Proc. ACM SIGACCESS Conf. Computers and Accessibility}, 2024. [Online]. Available: \url{https://arxiv.org/abs/2410.03022}

\bibitem{nganji2018pdf}
J. T. Nganji, ``An assessment of the accessibility of PDF versions of selected journal articles published in a WCAG 2.0 era (2014-2018),'' \emph{Learned Publishing}, vol. 31, no. 4, p. 391, 2018, doi: 10.1002/leap.1197.

\bibitem{paliwal2024forma11y}
S. Paliwal, J. Hoeflich, J. B. Jordan, R. Jain, V. I. Morariu, A. Siu, and J. Lazar, ``FormA11y - A tool for remediating PDF forms for accessibility: Designing a PDF Form Remediation Tool,'' \emph{ACM Trans. Comput.-Hum. Interact.}, vol. 32, no. 1, pp. 1--39, 2024, doi: 10.1145/3702317.

\bibitem{pdfa2021matterhorn}
PDF Association, ``The Matterhorn Protocol 1.1,'' 2021. [Online]. Available: \url{https://pdfa.org/resource/the-matterhorn-protocol/}

\bibitem{ribera2020proceedings}
M. Ribera, R. Pozzobon, and S. Sayago, ``Publishing accessible proceedings: the DSAI 2016 case study,'' \emph{Universal Access in the Information Society}, vol. 19, pp. 557--569, 2020, doi: 10.1007/s10209-019-00660-3.

\bibitem{silva2022android}
H. N. da Silva, S. R. Vergilio, and A. T. Endo, ``Accessibility Mutation Testing of Android Applications,'' \emph{Journal of Software Engineering Research and Development}, vol. 10, 2022, doi: 10.5753/jserd.2022.2133.

\bibitem{tafreshipour2024ma11y}
M. Tafreshipour, A. Deshpande, F. Mehralian, I. Ahmed, and S. Malek, ``Ma11y: A Mutation Framework for Web Accessibility Testing,'' in \emph{ISSTA 2024}, pp. 100--111, 2024, doi: 10.1145/3650212.3652113.

\bibitem{verapdfvalidation}
veraPDF Consortium, ``veraPDF Validation Documentation,'' 2026. [Online]. Available: \url{https://docs.verapdf.org/validation/}

\end{thebibliography}
\end{document}